\documentclass[prd,a4paper,preprint,preprintnumbers,nofootinbib,superscriptaddress]{revtex4-2}

\usepackage[a4paper, hdivide={1.91cm,,1.165cm}, vdivide={1.83cm,,3.0cm}]{geometry}

\usepackage{amsmath}
\usepackage{amsfonts}
\usepackage{graphicx}
\usepackage{xspace}
\usepackage{color}
\usepackage{units}
\usepackage{slashed}
\usepackage[hyperfootnotes=false]{hyperref}
\hypersetup{linkcolor=red,
citecolor=green,
filecolor=cyan,
urlcolor=magenta}

\usepackage{gensymb}
\usepackage{tabularx}
\usepackage{booktabs}
\usepackage{multirow,makecell}
\usepackage{xcolor}
\usepackage{setspace}
\usepackage{orcidlink}

\newcommand{\beq}{\begin{eqnarray}}
\newcommand{\eeq}{\end{eqnarray}}

\def\beq{\begin{equation}}
\def\eeq{\end{equation}}
\def\bea{\begin{eqnarray}}
\def\eea{\end{eqnarray}}
\def\beeq{\begin{eqnarray}}
\def\eeeq{\end{eqnarray}}
\def\ve{\vert}
\def\vel{\left|}
\def\ver{\right|}
\def\nnb{\nonumber}

\def\lla{\left<}
\def\rra{\right>}

\def\lrar{\leftrightarrow}  
\def\nnb{\nonumber}

\def\ba{\begin{array}}
\def\ea{\end{array}}

\def\xis0{{\Xi^{*0}}}

\def\g5{\gamma_5}

\def\es{&=&}

\def\ar{&+& \!\!\!}
\def\ek{&-& \!\!\!}

\def\cp{&\times& \!\!\!}

\def\olra{\stackrel{\leftrightarrow}}
\def\ola{\stackrel{\leftarrow}}
\def\ora{\stackrel{\rightarrow}}

\allowdisplaybreaks[1]

\begin{document}


\title{Looking for heavy axial tensor mesons via their strong decays in light cone QCD}

\author{T.~M.~Aliev\,\orcidlink{0000-0001-8400-7370}}
\email{taliev@metu.edu.tr}
\affiliation{Department of Physics, Middle East Technical University, Ankara, 06800, Turkey}

\author{S.~Bilmis\,\orcidlink{0000-0002-0830-8873}}
\email{sbilmis@metu.edu.tr}
\affiliation{Department of Physics, Middle East Technical University, Ankara, 06800, Turkey}
\affiliation{TUBITAK ULAKBIM, Ankara, 06510, Turkey}

\author{R.~A.~Ciftci\,\orcidlink{0000-0002-8017-1621}}
\email{ruken.ciftci@metu.edu.tr}
\affiliation{Department of Physics, Middle East Technical University, Ankara, 06800, Turkey}

\author{M.~Savci\,\orcidlink{0000-0002-6221-4595}}
\email{savci@metu.edu.tr}
\affiliation{Department of Physics, Middle East Technical University, Ankara, 06800, Turkey}

\date{\today}

\begin{abstract}
  In this study, the strong coupling constants of the unobserved heavy axial tensor mesons to the heavy vector and pseudoscalar $\pi$ and $K$ mesons, $D_2  D^* \pi$, $D_{s2}  D^* K$, $B_2  B^* \pi$, $B_{s2}  B^* K$  as well as  a heavy axial tensor to axial vector and light pseudoscalar $\pi$ and $K$-mesons, $D_2  D_1 \pi$, $D_{s2}  D_1 K$, $B_2  B_1 \pi$, $B_{s2}  B_1 K$ vertices have been investigated within the light cone QCD sum rules method. Having obtained the strong coupling constants, we estimate the corresponding decay widths that will hopefully be verified in future experiments.
\end{abstract}

\maketitle

\newpage


\section{Introduction\label{intro}}
The conventional quark model has been very successful in the classification of the hadrons so far. Within this theory, the states are represented by $J^{PC}$ quantum numbers in which $P=(-1)^{L+1}$ and $C=(-1)^{L+S}$ where $L$ and $S$ represent the orbital angular momentum and total spin of the state, respectively. Hence, the model predicts the existence of hadrons with $J^{PC} = 0^{++}, 0^{-+}, 1^{+-}, 1^{--}, 1^{++}, 2^{++}, 2^{-+},2^{--}$ and so on. Many of these states have already been discovered~\cite{ParticleDataGroup:2020ssz}. For instance mesonic light nonets $\langle \bar{q} q \rangle$ with $J^{PC} = 0^{-+}$ (pseudoscalar), $1^{--}$ (vector), $1^{++}$ (axial vector) are all well established. Moreover,  the light tensor states with $J^{PC}=2^{++}$ and the nonets of pseudotensor mesons with $J^{PC} = 2^{-+}$ are also well known. However, except $K_2(1820)$ meson~\cite{Aston:1993qc}, the nonet of axial tensor mesons with $J^{PC} = 2^{--}$ have not been observed yet.

A similar situation is present in the heavy sector as well. For example, the masses and the widths of the heavy mesons with $J^P = 2^+$; such as ${D}_2 (2420)$, ${D}_{s_2}(2573)$, ${B}_2(5747)$, and ${B}_{s_2}(5840)$ are observed via their strong decays such as ${D}_2 (2420) \to {D}^{+ \ast } \pi^-$, ${D}^+\pi^-$, ${D}^0\pi^+$, ${D}_{s_2}^+(2573) \to {D}^0 K^+$, ${B}_2 (5747) \to {B}^{+ \ast} \pi^-$, ${B}^+\pi^-$, ${B}_{s_2}^+(5840) \to {B}^+ K^-$ ~\cite{ParticleDataGroup:2020ssz}. These observations have triggered theoretical researches in studying the properties of the tensor mesons. Within the 3-point QCD sum rules method, the decay constants of the  ${D}_2 (2420)$ and ${D}_{s_2}^+(2573)$ mesons~\cite{Azizi:2014yua} and ${D}_2 (2420) \to {D} \pi$, ${B}_2 (5747) \to {B} \pi$, ${D}_{s_2}(2573) \to {D} K$, and ${B}_{s_2}(5840) \to {B} K$ transitions are analyzed~\cite{Wang:2014oca,Li:2015xka}. The light cone sum rules (LCSR) are also used to calculate the strong decay constants of these decays~\cite{Alhendi:2015rka}.

However, the states with $J^P = 2^-$ have still not been discovered except $D_2(2740)$~\cite{LHCb:2019juy}. Hence, investigating the properties of these hadrons anticipated by the quark model via analyzing their strong decays has vital importance.

The decays of axial tensor mesons $(T)$ into a vector $(V)$ and pseudoscalar $(\cal{P})$ ($T-V-\mathcal{P}$) as well as axial vector $(A)$ and pseudoscalar pairs ($T-A-\mathcal{P}$) are promising strong decay modes. In this paper, the strong decays of the axial tensor mesons,  $T-V-\mathcal{P}$ ($D_2  D^* \pi$, $D_{s2}  D^* K$, $B_2  B^* \pi$, $B_{s2}  B^* K$) and $T-A-\mathcal{P}$ ($D_2  D_1 \pi$, $D_{s2}  D_1 K$, $B_2  B_1 \pi$, $B_{s2}  B_1 K$) are investigated within the context of the light cone sum rules. 

The paper is organized as follows: In section~\ref{sec:2}, the light cone QCD sum rules are derived for the transitions of the axial tensor mesons to the heavy vector and light pseudoscalar $(\pi$ or $K)$ and to the heavy axial vector and light pseudoscalar mesons. Then, in section~\ref{sec:3}, we present the numerical analysis for the determination of the strong decay constants. The final section contains our concluding remarks. The expressions of distribution amplitudes and correlation functions are presented in the Appendix for brevity.
\section{Light cone QCD sum rules for the strong coupling constants
  of the heavy axial tensor mesons}
\label{sec:2}
In this section, we calculate the axial tensor-heavy vector-light pseudoscalar and  axial tensor-heavy axial vector-light pseudoscalar vertices. For this purpose, we start by considering the following correlation function,
\begin{equation}
  \label{ejly01}
\Pi_{\alpha \beta \tau} (p,q) = i \int d^4x e^{ipx} \lla {\cal P}(q) \vel
J_{\alpha \beta} (x) J_\tau^{\dag V(A)} (0) \ver 0 \rra~,  
\end{equation}
where $J_{\alpha \beta}$ and $J_\tau^{V(A)}$ are the interpolating currents of the heavy axial tensor meson and the heavy vector (axial) mesons, respectively
\begin{equation}
  \label{eq:current}
  \begin{split}
J_{\alpha \beta} (x)                 & = {1\over 2} \Big[ \bar{q} (x) \gamma_\alpha \gamma_5 \olra{\cal D}_\beta Q (x) +
                 \bar{q} (x) \gamma_\beta \gamma_5 \olra{\cal D}_\alpha Q(x)   \Big]~, \\
                 J_\tau^{V(A)} & = \bar{q} \gamma_\tau (\gamma_5) Q~.
  \end{split}
\end{equation}
The covariant derivative $\olra{\cal D}_\alpha$ is defined as,
\begin{equation}
  \label{eq:covariant}
  \begin{split}
\olra{\cal D}_\alpha  &= {1\over 2} \Big( \ora{\cal D}_\alpha  - \ola{\cal D}_\alpha  \Big)~, \\
\overrightarrow{\cal D}_\alpha  &= \overrightarrow{\partial}_\alpha  -  i {g\over 2} \lambda^a G_\alpha^a  ~, \\
\overleftarrow{\cal D}_\alpha  &= \overleftarrow{\partial}_\alpha  +  i {g\over 2} \lambda^a G_\alpha^a  ~, 
  \end{split}
\end{equation}
in which $\lambda^a$ are the Gell-Mann matrices, $g$ is the coupling constant, and $G_\alpha^a$ is the external gluon field.

To derive the LCSR for the relevant strong coupling constants, the correlation function is calculated both in terms of the hadrons and in terms of the quarks and gluons. Then, by matching the coefficients of the corresponding Lorentz structures, the desired LCSR can be obtained .

The hadronic representation of the correlation function can be obtained by inserting a complete set of intermediate hadrons having the same quantum numbers with the interpolating current into the correlation function. And isolating the ground state contributions from the axial tensor and heavy vector (axial) mesons, we obtain the correlation function as follows.
\bea
\label{ejly02}
\Pi_{\alpha \beta \tau} = {\langle {\cal P} (q) T(p) \ve V(A)\rangle
\langle 0 \ve J_{\alpha \beta} \ve T \rangle
\langle V(A) J_\tau^{\dag V(A)} \ve 0 \rangle
\over (p^2-{m_T}^2) ( p^{\prime 2} - m_{V(A)}^2)}+ \cdots ~.
\eea
The matrix elements in Eq. (\ref{ejly02}) are defined as;
\begin{equation}
  \label{ejly03}
  \begin{split}
\lla 0 \vel J_{\alpha \beta} \ver T(p) \rra &= f_T {m_T}^3 \varepsilon_{\alpha \beta} (p)~, \\
\lla 0 \vel J_\tau^{V(A)} \ver V(A) \rra &= f_{V(A)} m_{V(A)} \varepsilon_\tau^{V(A)}(p^\prime)~, \\
\lla {\cal P}(q) T(p) \ve V\rra &= g_{TV\mathcal{P}} \varepsilon^{\rho \sigma} q_{\rho} \varepsilon_\sigma^V~, \\
\lla {\cal P}(q) T(p) \ve A\rra &= g_{TA\mathcal{P}} \varepsilon^{\rho \sigma \eta \lambda} p_\rho \varepsilon_{\sigma \varphi} q^\varphi p_\eta^\prime \varepsilon_\lambda^A (p^\prime)~,    
  \end{split}
\end{equation}
where $\varepsilon_{\alpha \beta}$, $\varepsilon^{V(A)}$ are the polarizations and $f_T$, $f_{V(A)}$ are the decay constants of the tensor and vector (axial) mesons, respectively, and $q_\alpha$ is the momentum of the pseudoscalar mesons. $g_{TV\mathcal{P}}$ and $g_{TA\mathcal{P}}$ are the strong coupling constants of the corresponding interactions.

Summation over the polarization of the tensor and vector (axial) mesons is performed in accordance with the following formulas, 
\begin{equation}
  \label{ejly04}
  \begin{split}
    \sum_{spins} \varepsilon_{\alpha\beta} (p) \varepsilon_{\rho\sigma}^\ast (p) &=  {1\over 2} \Big[ \widetilde{g}_{\alpha\rho} \widetilde{g}_{\beta\sigma} + \widetilde{g}_{\alpha\sigma} \widetilde{g}_{\beta\rho}\Big] -
{1\over 3} \widetilde{g}_{\alpha\beta} \widetilde{g}_{\rho\sigma}~, \\
\sum_{spins} \varepsilon_\alpha(p^\prime)  \varepsilon_\beta^\ast (p^\prime) &= \widetilde{g}_{\alpha\beta} (p^\prime)
  \end{split}
\end{equation}
where $\widetilde{g}_{\alpha\beta} (p) = g_{\alpha\beta} - {p_\alpha p_\beta \over p^2}$.

Using the definitions of the matrix elements given in Eq.(\ref{ejly03}), and performing summation over the polarizations of the tensor and vector (axial) mesons via Eq.\eqref{ejly04}, we obtain the following expressions of the correlation functions from the hadronic side.
\bea
\label{ejly05}
\Pi_{\alpha\beta\tau}^A\es {f_{T} {m_T}^3 f_A m_A g_{TA\mathcal{P}} \over 2 (p^2-{m_T}^2)
(p^{\prime 2}-m_A^2)} \epsilon_{\alpha\lambda\varphi\tau}  \Bigg\{
 p^\lambda q^\varphi q_\beta -
{{m_T}^2-m_A^2+q^2 \over 2 {m_T}^2} p^\lambda q^\varphi p_\beta \Bigg\}
+ (\alpha \lrar \beta)~, \\ \nnb \\
\label{ejly06}
\Pi_{\alpha\beta\tau}^V \es {f_{T} {m_T}^3 f_V m_V g_{TV\mathcal{P}} \over 2 (p^2-{m_T}^2)
(p^{\prime 2}-m_V^2)} \Bigg\{
{{m_T}^2-m_V^2+q^2 \over 2 {m_T}^2}
(p_\alpha g_{\beta\tau} + p_\beta g_{\alpha\tau}) \nnb \\
\ar {m_V^4 - ({m_T}^2 - q^2)^2 \over 6 {m_T}^2 m_V^2}
p_\tau g_{\alpha\beta}
- (q_\alpha g_{\beta\tau} + q_\beta g_{\alpha\tau}) +
{({m_T}^2+m_V^2-q^2)^2 \over 6 {m_T}^2 m_V^2} q_\tau g_{\alpha\beta} \nnb \\
\ek {{m_T}^4 - m_V^4 +
q^2 (4 {m_T}^2 + q^2) \over 3 {m_T}^4  m_V^2} p_\alpha p_\beta p_\tau
+ {2 \over m_V^2} q_\alpha q_\beta q_\tau \nnb \\
\ek  {{m_T}^2-m_V^2+q^2 \over {m_T}^2 m_V^2}
(p_\alpha q_\beta q_\tau +  p_\beta q_\alpha q_\tau)
- {2 \over m_V^2} p_\tau q_\alpha q_\beta \nnb \\
\ar {{m_T}^4 + (m_V^2 - q^2)^2 -
4 {m_T}^2 (m_V^2 - q^2) \over 3 {m_T}^4 m_V^2} p_\alpha p_\beta q_\tau
+ { {m_T}^2 + q^2 \over  {m_T}^2 m_V^2} (p_\alpha p_\tau q_\beta + p_\beta p_\tau
q_\alpha) \Bigg\}~.
\eea
As a next step, we calculate the correlation function from the QCD side. After contracting the
heavy quark fields using Wick's theorem, the correlation function can be written as,
\bea
\label{ejly07}
\Pi_{\alpha \beta \tau} = -{1\over 2} \int d^4x e^{ipx} \lla {\cal P}(q) \vel
\bar{q}(x) (\gamma_\alpha \gamma_5) \olra{\cal D}_\beta (x) S_Q (x) \gamma_\tau 
(\gamma_\beta \gamma_5) q(0) 
+ (\alpha \lrar \beta) \ver 0 \rra~,
\eea
where $S_Q (x)$ is the heavy quark propagator whose expression in the
coordinate representation is given as,
\bea
\label{ejly08}
S_Q(x) \es  {m_Q^2 \over 4 \pi^2} \Bigg[ {K_1(m_Q\sqrt{-x^2}) \over
\sqrt{-x^2}} +  {\slashed{x} \over
\left(\sqrt{-x^2}\right)^2} K_2(m_Q\sqrt{-x^2}) \Bigg] \nnb \\
\ek { i g_s m_Q \over 16 \pi^2} \int_0^1 du G_{\mu\nu}(ux) \Bigg[
\sigma^{\mu\nu} K_0(m_Q\sqrt{-x^2}) +
 \left( u \sigma^{\mu\nu} \slashed{x} +
\bar{u} \slashed{x} \sigma^{\mu\nu}\right) {K_1 (m_Q\sqrt{-x^2})\over \sqrt{-x^2}}
 \Bigg]~,
\eea
where $K_1$ and $K_2$ are the modified Bessel function of the first and second kind and $\bar{u} = 1-u$.
It should be noted that we neglect the contributions of the four-particle operators such as $\bar{q} G^2 q$ and $\bar{q}q \bar{q}q$ in our calculations since we anticipate that they will be small (see for example \cite{BALITSKY1989541}).

The theoretical part of the correlation function is calculated by using the operator product expansion (OPE) in the deep-Euclidean region $p^2 \to - \infty$. It follows from Eq.\eqref{ejly01} that it is necessary to know the matrix elements  $\lla {\cal P}(q) \vel \bar{q}(x) \Gamma q^\prime (0) \ver 0 \rra$ of the non-local operators between the vacuum and the light pseudoscalar
mesons where $\Gamma$ is one of the members of the full-set of the Dirac matrices. These matrix elements are represented in terms of the distribution amplitudes (DAs) of the pseudoscalar mesons~\cite{Ball:2006wn,Ball:2004ye} and presented in Appendix \ref{appendix_das} for completeness. With these DAs, the theoretical part of the correlation functions can be calculated.

Having obtained the correlation function from both the hadronic and theoretical sides, by matching the coefficients of the corresponding Lorentz structures, and performing the Borel transformation over the variables $-p^2$ and $-p^{\prime 2}$, which suppresses the higher states and continuum contributions of the desired sum rules, we obtain the sum rules for the strong decay constants of the heavy axial tensor mesons to vector (axial) and pseudoscalar meson transitions as follows.
\bea
\label{ejly12}
 g_{TA\mathcal{P}} e^{-{m_T}^2/M_1^2 - m_A^2/M_2^2} A_i \es 
\Pi_i^A \nnb \\
  g_{TV\mathcal{P}} e^{-{m_T}^2/M_1^2 - m_V^2/M_2^2} B_i \es
\Pi_i^V~,  
\eea
where $A_i$ and $B_i$ are the relevant coefficients of the structures present in the $T-A-{\cal P}$ $(T-V-{\cal P})$ transitions [see Eqs.\eqref{ejly05} and \eqref{ejly06}]. The explicit expressions of $\Pi_i^A$, $\Pi_i^V$, and  $A_i$, $B_i$ are presented in Appendix~\ref{appendix_a}.
\section{Numerical analysis}
\label{sec:3}
In this section, we calculate the strong coupling constants of the axial tensor mesons with  heavy vector (axial) and light pseudoscalar mesons using the LCSR results derived in the previous section.
%
\begin{table}[hbt]
\renewcommand{\arraystretch}{1.4}
\addtolength{\arraycolsep}{1.6pt}
\small
\centering
  \begin{tabular}{cccccc}
    \toprule
    \multirow{2}{*}{\shortstack{quark mass \\ $(\rm{GeV})$}} \cite{ParticleDataGroup:2020ssz}                                    & \multicolumn{4}{c}{meson mass $(\rm{GeV})$}       &
                                                                                                                                                                        \multirow{2}{*}{\shortstack{decay constant \\ $(\rm{GeV})$}}~\cite{Wang:2015mxa}                           \\
    \cmidrule(lr){2-5}
                                                                                & $T$~\cite{Chen:2011qu}       &  $V$~\cite{ParticleDataGroup:2020ssz}  & A~\cite{ParticleDataGroup:2020ssz} & $\mathcal{P}$~\cite{ParticleDataGroup:2020ssz}  \\   
    \cmidrule(lr){1-1} \cmidrule(lr){2-2} \cmidrule(lr){3-3} \cmidrule(lr){4-4} \cmidrule(lr){5-5} \cmidrule(lr){6-6}

    $m_c= (1.275 \pm 0.025)$ & $m_{D_2} = 2.74$\cite{ParticleDataGroup:2020ssz}              & $m_{D^{*}}= 2.01$ & $m_{D_{1}}= 2.42$ & $m_\pi = 0.140$ & $f_{{D}^\ast}=(0.263 \pm 0.021)$  \\
    $m_b= (4.18 \pm 0.03)$   & $m_{D_{s_2}} =(3.01\pm 0.21)$   & $m_{B^{*}}= 5.32$   & $m_{B_{1}}= 5.72$ & $m_K = 0.494$    & $f_{{D}_1}=(0.332 \pm 0.018)$  \\
                             & $m_{B_2} = (5.66\pm 0.33)$      &                   &                   &             & $f_{{B}^\ast}=(0.213 \pm 0.018) $ \\
                             & $m_{B_{s_2}}= (6.40 \pm 0.25)$   &                    &                   &              & $f_{{B}_1}=(0.335 \pm 0.018)$     \\
\bottomrule
  \end{tabular}
  \caption{The values of the input parameters needed for the numerical calculations are presented.}
  \label{tab:input_parameters}
\end{table}


The values of the input parameters used in the numerical calculations are shown in Table~\ref{tab:input_parameters}. For the mass of the axial tensor mesons except for the observed $D_2$ meson we used the sum rule predictions obtained in~\cite{Chen:2011qu}. The distribution amplitudes are the major input parameters for the light-cone sum rules. In our case, we require the DAs of pseudoscalar $\pi$ and $K$ mesons. For the sake of completeness, we include these definitions in Appendix~\ref{twist-four}, which are provided in~\cite{Ball:2006wn,Ball:2004ye}.

\begin{figure}[!]
\includegraphics[width=0.45\textwidth]{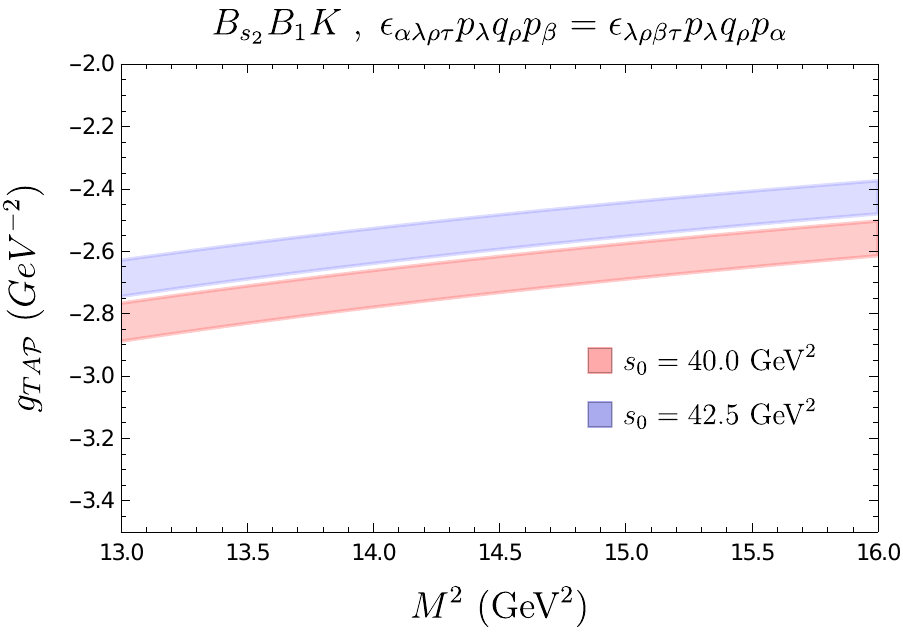}
\includegraphics[width=0.45\textwidth]{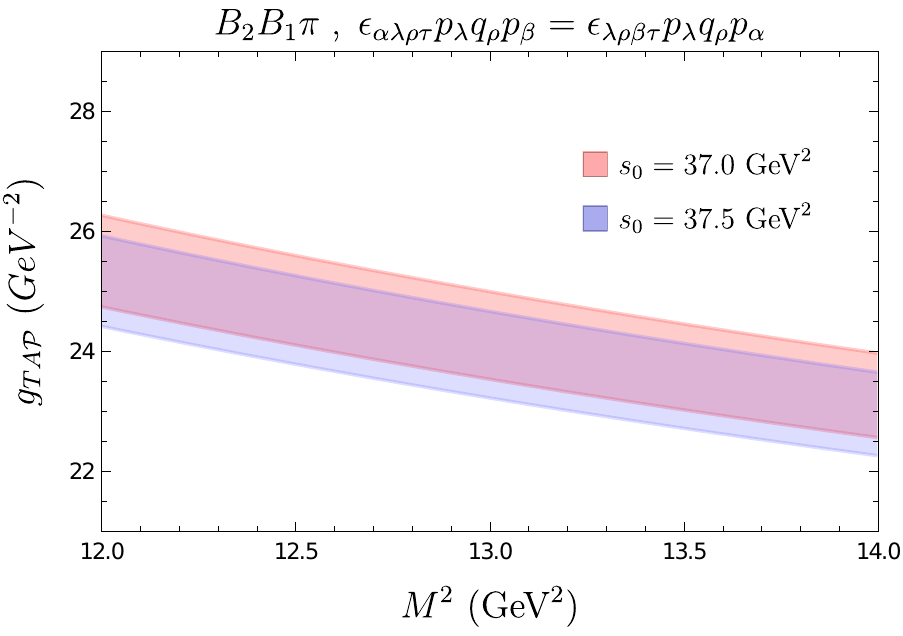} \\
\includegraphics[width=0.45\textwidth]{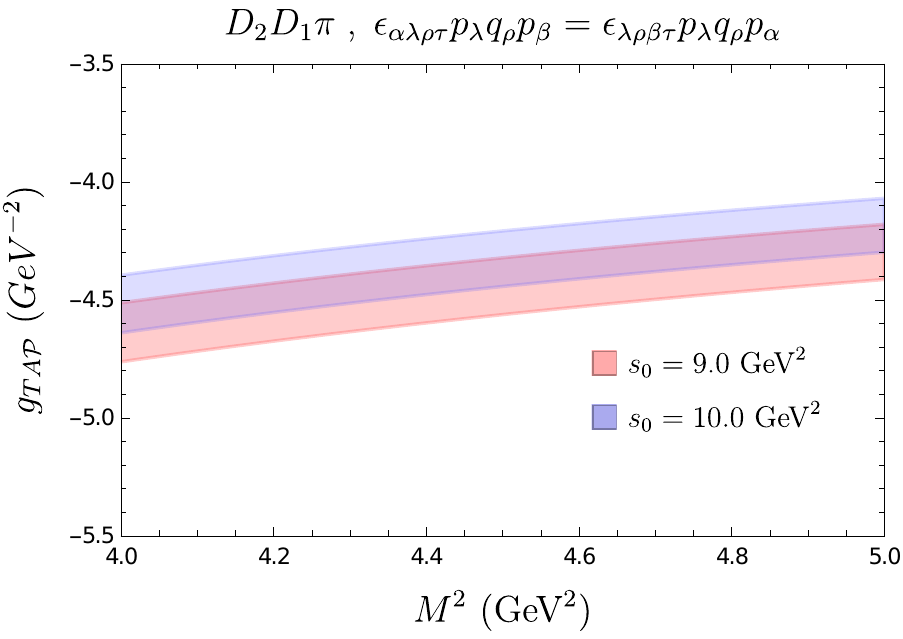}
\includegraphics[width=0.45\textwidth]{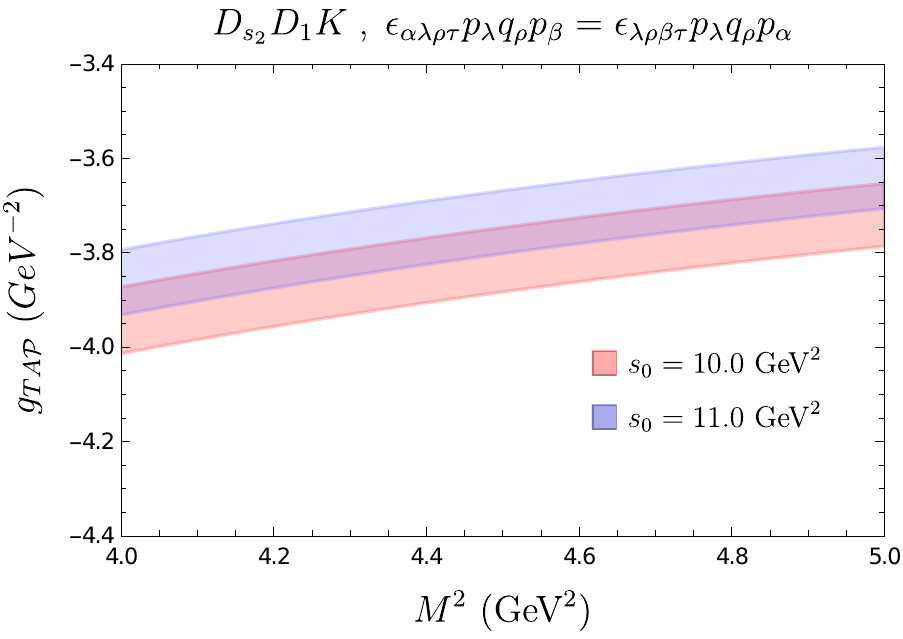} \\
\includegraphics[width=0.45\textwidth]{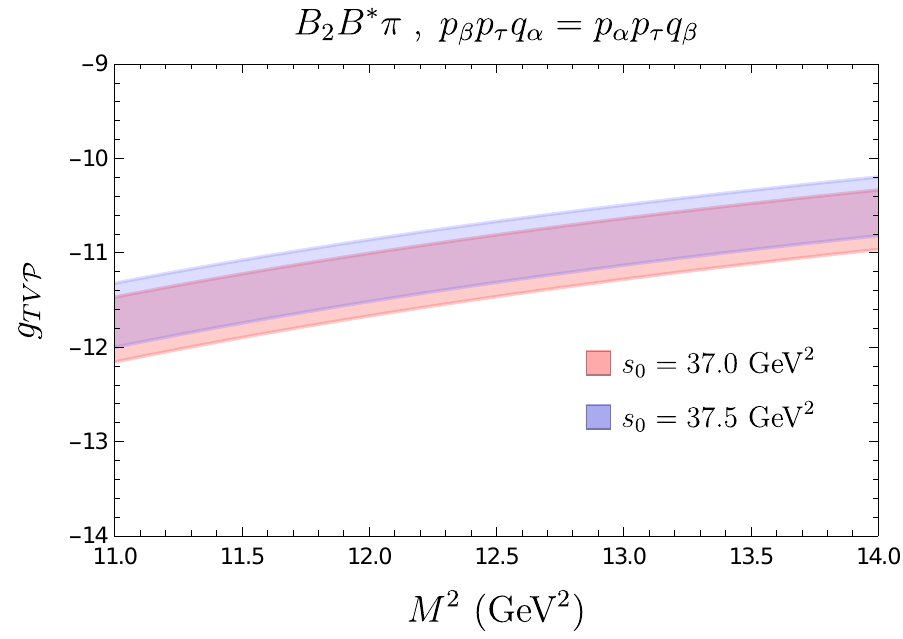}
\includegraphics[width=0.45\textwidth]{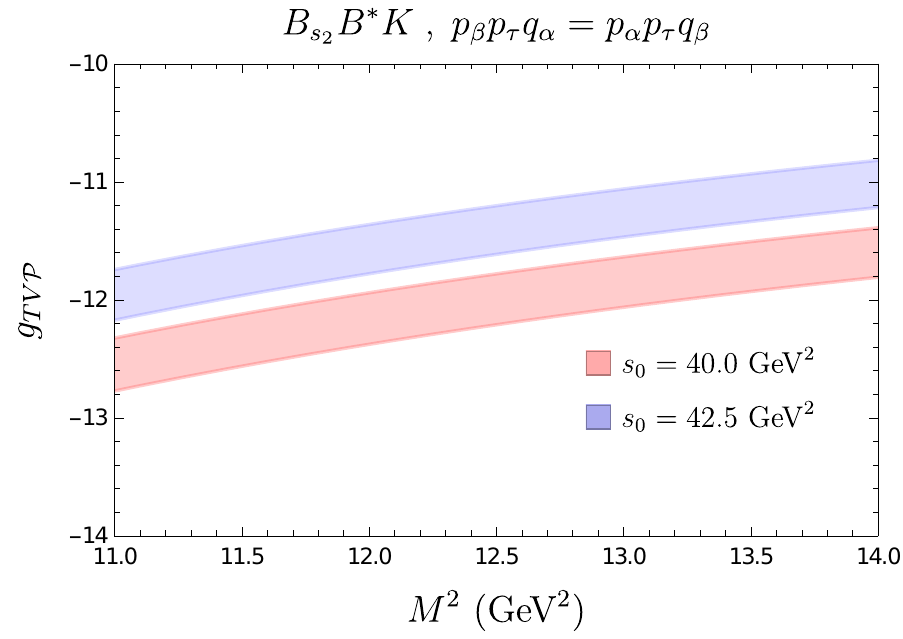} \\
\includegraphics[width=0.45\textwidth]{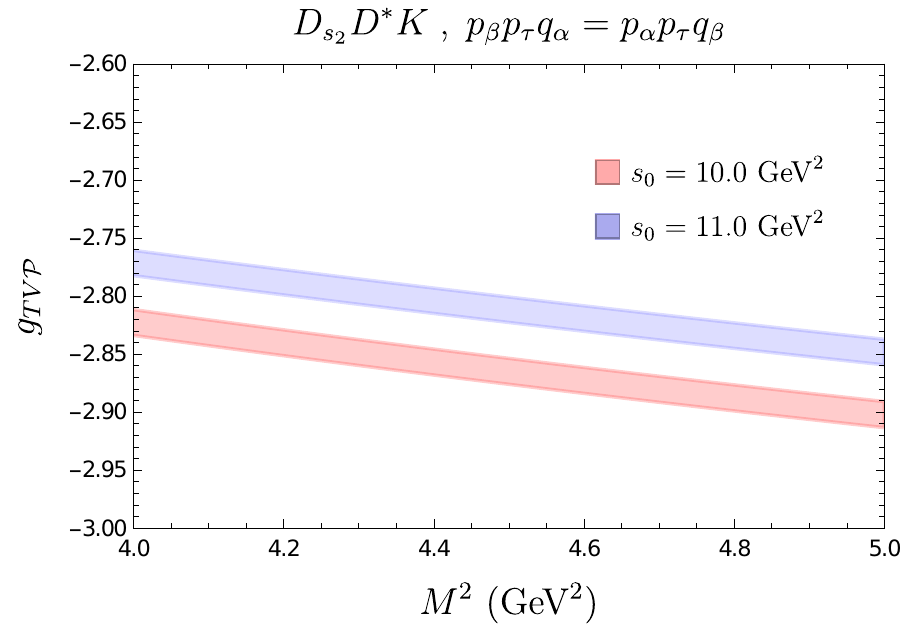}
\includegraphics[width=0.45\textwidth]{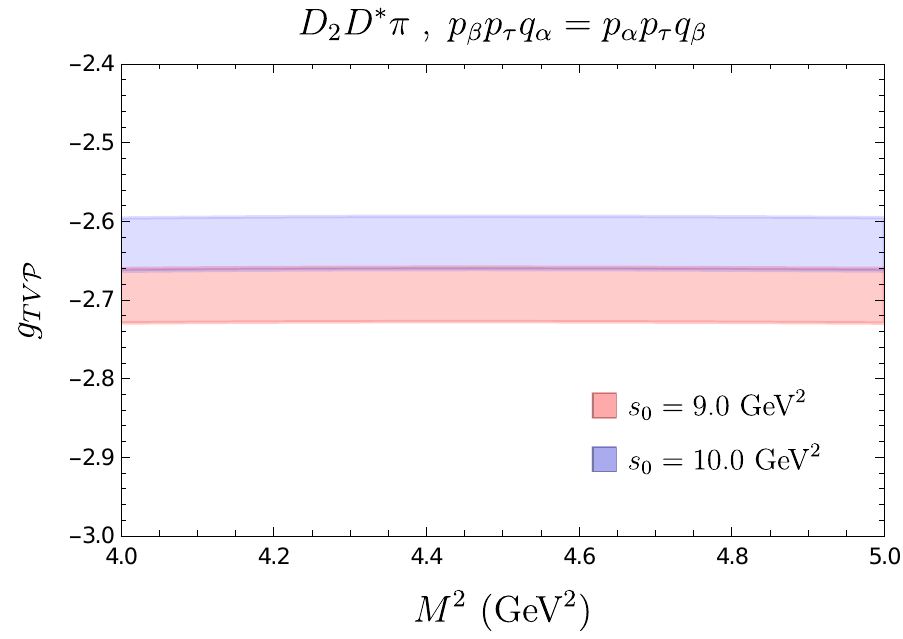} \\
 \caption{
The variation of the $T-V-P$ and $T-A-P$ vertices on the Borel mass parameter at the fixed values of $s_0$ is shown for the considered Lorentz structures. The shaded regions show uncertainty due to the first and second momenta of DAs of pseudoscalar mesons.
 }
 \label{fig:fig1}
 \end{figure}

Note that the DAs of pseudoscalar mesons involve the first and second momenta $a_1^P$ and $a_2^P$, respectively. However, there is $30 \%$ uncertainty in these parameters' determination in QCD sum rules~\cite{Ball:1998je}. As a result, the choice of these moments has a significant impact on the determination of the coupling constants~\cite{Alhendi:2015rka}.  Numerous research have been conducted recently~\cite{Donnellan:2007xr,Braun:2015axa,Arthur:2010xf,Braun:2006dg,DelDebbio:2002mq} to compute these parameters inside lattice QCD, and they have now been more precisely defined~\cite{RQCD:2019osh} as shown in Table~\ref{tab:daparams}. The uncertainties caused by these factors are taken into consideration in our calculations. (See Figure~\ref{fig:fig1} (color online)).

\begin{table}[hbt]
 \renewcommand{\arraystretch}{1.2}
   \setlength{\tabcolsep}{12pt}
  \begin{tabular}{ccc}
    \toprule
                       & $\pi$ & $K$   \\
    \midrule
$a_1^{\cal P}$          & 0     & $0.0525^{+31}_{-33}$ \\
$a_2^{\cal P}$  & $ 0.116^{+19}_{-20}$  & $0.106^{+15}_{-16}$   \\
    \bottomrule
  \end{tabular}
  \caption{The values of the parameters of the wave function in DAs obtained within lattice QCD calculations ~\cite{RQCD:2019osh} (at the renormalization scale $\mu = 2~\rm{GeV}$) } 
  \label{tab:daparams}
\end{table}

The sum rules also include two auxiliary parameters, i.e., the continuum threshold $s_0$ and the Borel mass parameters $M_1^2$ and $M_2^2$. Note that the results would be independent of the Borel mass parameters if it were possible to calculate the correlation function exactly. Here we employ the following approach for the Borel mass parameters. Due to the fact that the two momenta $p^2$ and $p^{\prime 2}$, and therefore the corresponding two Borel parameters $M_1^2$ and $M_2^2$ are independent variables, we account for the mass difference between heavy axial tensor and heavy vector (axial) mesons. By setting the Borel parameters to a fixed value, the sum rule can be enhanced to some extent.
\begin{equation}
  \label{eq:ratio}
  \frac{M_1^2}{M_2^2} = \frac{m_T^2}{m_{V(A)}^2}~.
\end{equation}
With this equation and using the definition $\frac{1}{M_1^2} + \frac{1}{M_2^2} = \frac{1}{M^2}$ we get,
\begin{equation}
  \label{eq:borel}
  \frac{m_T^2}{M_1^2} + \frac{m_{V(A)}^2}{M_2^2} \to \frac{\mu^2}{M^2}~,
\end{equation}
where $\mu^2 = \frac{2 m_T^2 m_{V(A)}^2}{m_T^2 + m_{V(A)}^2}$. Hence, instead of two, we can work with one Borel mass parameter, i.e., $M^2$, and Eq.~\eqref{ejly12} takes the following form.
\begin{equation}
  \label{ejly12a}
  \begin{split}
  g_{TA\mathcal{P}} e^{-\mu^2/M^2 } A_i &= \Pi_i^A  \\
 g_{TV\mathcal{P}} e^{-\mu^2/M^2 } B_i &= \Pi_i^V~.      
  \end{split}
\end{equation}
By imposing the requirement that the contributions resulting from higher states continuum should account for less than $30\%$ of the final result, the operational region of the Borel mass parameter is established. The lower bound for $M^2$ can be obtained by ensuring that the highest twist term's contribution is $15\%$ smaller than the contributions from the leading twist term. These two conditions allow us to determine the working region of $M^2$. Besides, the working region of $s_0$ is determined from the analysis of two-point sum rules~\cite{Chen:2011qu}. We present the working regions of $M^2$ and $s_0$ in Table~\ref{tab:region}. Obviously, physical quantities should almost be independent of the variation of these auxiliary parameters in their working regions. In Figure~\ref{fig:fig1}, we exhibit the dependency of strong coupling constants for the considered vertices with respect to the variation of $M^2$ for the chosen Lorentz structure at the fixed values of continuum threshold $s_0$. The plots show the weak dependency on the variation of the parameters and verify the goodness of the chosen working region. 
%
%
\begin{table}[hbt]
\renewcommand{\arraystretch}{1.2}
\setlength{\tabcolsep}{12pt}
  \begin{tabular}{lcc}
    \toprule
                         &     $ M^2(\rm{GeV^2})$            & $  s_0(\rm{GeV^2})$           \\
    \midrule
$g_{D_2   D^* \pi}$     &   $  4 - 5 $  & $ 9 - 10$  \\ 
$g_{D_2   D_1 \pi}$     &   $  4 - 5 $  & $ 9 - 10$  \\ 
$g_{D_{s_2}   D^* K} $  &   $ 4 - 5$    & $ 10 - 11$  \\ 
$g_{D_{s_2}   D_1 K} $  &   $ 4 - 5$    & $ 10 - 11$  \\ 
$g_{B_2   B^* \pi}$     &   $  11 - 14 $  & $ 37 - 37.5$  \\ 
$g_{B_2   B_1 \pi}$     &   $  12 - 14 $  & $ 37 - 37.5$  \\ 
$g_{B_{s_2}   B^* K} $  &   $ 11 - 14$    & $ 40 - 42.5$  \\ 
$g_{B_{s_2}   B_1 K} $  &   $ 13 - 16$    & $ 40 - 42.5$  \\ 
    \bottomrule
  \end{tabular}
  \caption{Working regions of the Borel mass parameter $M^2$ and the continuum threshold $s_0$ for the considered vertices.}
  \label{tab:region}
\end{table}

Having determined the domains of $M^2$ and $s_0$, we are ready to calculate the relevant strong coupling constants of the $T-V-\mathcal{P}$ and $T-A-\mathcal{P}$ interactions. The numerical results of the strong coupling constants for all possible Lorentz structures are presented in Tables~\ref{tab:tabB2} and \ref{tab:8}. The uncertainties due to the first and second momenta of pseudoscalar meson DAs as well as  errors coming from input parameters and variation of $s_0$ and $M^2$ are taken into account. Our findings show that the strong coupling constants depend heavily on the choice of Lorentz structures, and the numerical values vary through a wide range between $0.2$ and $100.0$. A similar outcome was observed for the strong coupling constants of the heavy tensor mesons with vector and pseudoscalar heavy mesons in participation of the light pseudoscalar mesons~\cite{Alhendi:2015rka}.

Our numerical analysis shows that the best convergence of OPE is achieved for the  $p_\beta p_\tau q_\alpha$ ($p_\alpha p_\tau q_\beta$) structure. Besides, once we consider the heavy quark and chiral symmetry then the coupling constants for $D_2  D^* \pi$, $D_{s_2}  D^* K$ and  $B_2  B^* \pi$, $B_{s_2}  B^* K$ vertices are expected to be close to each other. We see from Tables~\ref{tab:tabB2} that this condition is satisfied with good accuracy for the $p_\alpha p_\tau q_\beta$ ($p_\beta p_\tau q_\alpha$) structure. And notice that the reason for large difference between $B_2 B_1 \pi$ and $B_{s2} B_1 K $ couplings is due to the factor $ \frac{m_T^2 - m_A^2 + m_\mathcal{P}^2}{2 m_T^2}$ in Eq.\eqref{ejly05}. Hence, for further analysis, we will use the coupling constants corresponding to the Lorentz structures $p_\alpha p_\tau q_\beta$ for $T-V-\mathcal{P}$ and $\epsilon_{\alpha \lambda \rho \tau} p_{\lambda} q_\rho p_\beta$ for $T-A-\mathcal{P}$ interactions. For brevity, these values are also collected in Table~\ref{tab:results}.
\begin{table}[hbt]
   \renewcommand{\arraystretch}{1.2}
   \setlength{\tabcolsep}{12pt}
   \centering
  \begin{tabular}{ccccc}
    \toprule
    Structures                                          & $B_2    B^\ast \pi$ & $ B_{s_2}   B^\ast K$ & ${D_2}   { D}^\ast \pi$ & ${D_{s_2}}   {D}^\ast K$ \\
\midrule
    $g_{\beta \tau} p_\alpha = g_{\alpha \tau} p_\beta$    & $23.01 \pm 1.39$      & $8.50 \pm 1.53$        & $ 3.46  \pm 0.84 $        & $3.99 \pm 0.78$           \\
$ g_{\alpha\beta} p_\tau$                               & $-0.38 \pm 0.01$       & $-0.48 \pm 0.02$        & $-0.57 \pm 0.01$          & $-0.82 \pm 0.01$           \\
    $ g_{\beta\tau} q_\alpha = g_{\alpha\tau} q_\beta$    & $0.77 \pm 0.01$       & $1.00 \pm 0.13$         & $ 0.44 \pm 0.16$          & $0.42 \pm 0.18$            \\
$ g_{\alpha\beta} q_\tau$                               & $-5.78 \pm 0.25$      & $-7.15 \pm 0.13$        & $ -3.75 \pm 0.16$         & $-4.67 \pm 0.25$           \\
$ p_\alpha p_\beta p_\tau$                              & $-0.82 \pm 0.05$      & $-1.15 \pm 0.09$        & $ -0.80 \pm 0.04$         & $-1.02 \pm 0.05$           \\
$ q_\alpha q_\beta q_\tau$                              & $-5.10 \pm 0.24$      & $-6.90 \pm 0.32$        & $ -2.08 \pm 0.05$         & $-2.52 \pm 0.05$           \\
$ p_\alpha q_\beta q_\tau = p_\beta q_\alpha q_\tau$    & $106.60 \pm 5.16$     & $50.51 \pm 2.32$        & $ 8.52 \pm 0.21$         & $8.80 \pm 0.19$           \\
$p_\tau q_\alpha q_\beta$                               & $5.93 \pm 0.28$       & $7.04 \pm 0.35$         & $ 1.82 \pm 0.05$          & $2.10 \pm 0.03$            \\
$p_\alpha p_\beta q_\tau$                               & $23.49 \pm 1.25$      & $38.01 \pm 0.90$        & $ 15.18 \pm 0.43$         & $25.56 \pm 0.74$           \\
$ p_\beta p_\tau q_\alpha = p_\alpha p_\tau q_\beta$    & $-11.11 \pm 0.52$     & $-11.73 \pm 0.58$        & $ -2.66 \pm 0.07$        & $-2.85 \pm 0.05$           \\
   \bottomrule
   \end{tabular}
   \caption{The values of strong coupling constants for $T-V-\mathcal{P}$ vertices.}
   \label{tab:tabB2}
 \end{table}
%

\begin{table}[hbt]
   \renewcommand{\arraystretch}{1.1}
   \setlength{\tabcolsep}{6pt}
   \centering
  \begin{tabular}{ccccc}
    \toprule
    Structures                                                                                                    & ${D_2}  { D}_1 \pi$   & ${D_{s_2}}   {D}_1 K$ & ${B_2}   {B}_1 \pi$ & $ B_{s_2}   { B}_1 K$ \\
\midrule
$\epsilon_{\alpha \lambda \rho \tau} p^\lambda q^\rho p_\beta =  \epsilon_{\lambda \rho\beta\tau} p^\lambda q^\rho p_\alpha$    & $-4.31 \pm 0.20$         & $-3.74 \pm 0.13$        & $ 24.20  \pm 1.11$    & $-2.55 \pm 0.15$         \\
$ \epsilon_{\alpha \lambda \rho \tau} p^\lambda q^\rho q_\beta  =   \epsilon_{\lambda \rho \beta \tau} p^\lambda q^\rho q_\alpha$ & $0.36 \pm 0.02$          & $0.43 \pm 0.01$         & $ 0.13 \pm 0.01$      & $0.15 \pm 0.01$          \\
    \bottomrule
     \end{tabular}
   \caption{The values of  strong coupling constants (in units of $\rm{GeV^{-2}}$) for $T-A-\mathcal{P}$ vertices.}
   \label{tab:8}
 \end{table}



\begin{table}[hbt]
 \renewcommand{\arraystretch}{1.2}
   \setlength{\tabcolsep}{12pt}
  \begin{tabular}{lcc}
    \toprule
    Decays                  & $g_{TV\mathcal{P}}$                 & $\Gamma(\rm{GeV})$                 \\
    \midrule
${B_2} \to { B}^\ast \pi$   & $-11.11 \pm 0.52$                    & $(1.32 \pm 0.12) \times 10^{-3} $    \\ 
${B_{s_2}} \to { B}^\ast K$ & $-11.73 \pm 0.58$                    & $ (3.01 \pm 0.30) \times 10^{-2}$  \\ 
${D_2} \to { D}^\ast \pi$   & $-2.66 \pm 0.07$                    & $(3.14 \pm 0.17) \times 10^{-3}$   \\ 
${D_{s_2}} \to { D}^\ast K$ & $-2.85 \pm 0.05$                    & $(4.75 \pm 0.17) \times 10^{-3}$   \\ 
\midrule
    Decays                  & $g_{TA\mathcal{P}} (\rm{GeV^{-2})}$ & $\Gamma(\rm{GeV})$                 \\
    \midrule
    ${D_2} \to { D}_1 \pi$  & $-4.31 \pm 0.20$                    & $ (2.08 \pm 0.12) \times 10^{-4} $ \\ 
${D_{s_2}} \to { D}_1 K$    & $-3.74 \pm 0.13$                    & $(2.16 \pm 0.15) \times 10^{-4} $  \\ 
${B_2} \to { B}_1 \pi$      & $24.20 \pm 1.11$                    & NA                                 \\
${B_{s_2}} \to { B}_1 K$    & $-2.55 \pm 0.15$                    & $ (7.98 \pm 0.94 ) \times 10^{-4}$   \\
    \bottomrule
  \end{tabular}
  \caption{Strong coupling constants and decay widths $\Gamma$ of heavy axial tensor mesons.}
  \label{tab:results}
\end{table}

Using Eqs. (\ref{ejly03}) the widths for the decays under consideration can be calculated straightforwardly for $T V \mathcal{P}$ and $T A \mathcal{P}$, respectively as follows. 
\begin{equation}
  \label{decaywidth}
  \begin{split}
  \Gamma_{TV\mathcal{P}}   &= \frac{g_{TV\mathcal{P}}^2 \lambda \left(m_T^2,m_V^2,m_P^2\right){}^{3/2} \big(\lambda \left(m_T^2,m_V^2,m_P^2\right)+10 m_T^2 m_V^2\big)}{1920 \pi  m_T^7 m_V^2}~, \\
    \Gamma_{ T A \mathcal{P}} &= \frac{g_{TA\mathcal{P}}^2 \lambda \left(m_T^2,m_A^2,m_P^2\right){}^{5/2}}{1280 \pi  m_T^5}~,
  \end{split}
\end{equation}
where $\lambda(a,b,c) = a^2 + b^2 + c^2 - 2 a b - 2 a c - 2 b c$.

With the obtained strong coupling constants, the corresponding decay widths are calculated and the results are presented in Table~\ref{tab:results}. Recall that $B_2 \to B_1 \pi$ decay is not kinematically not allowed.

Our predictions demonstrate that even though the heavy axial tensor states anticipated by the quark model have huge decay widths, they may still be seen in existing and future planned experiments. We hope that these results will help to comprehend the nature of the prospective axial tensor mesons.
\section{Conclusion\label{conclusion}}
In this study, using the light cone sum rules approach, we analyzed the strong couplings of the potential heavy axial tensor mesons to vector (axial) and pseudoscalar $\pi$ and $K$ mesons. Specifically, we studied $D_2  D^* \pi$, $D_{s2}  D^* K$, $B_2  B^* \pi$, $B_{s2}  B^* K$ $(T-V\mathcal{P}$)  and $D_2  D_1 \pi$, $D_{s2}  D_1 K$ $B_2  B_1 \pi$, $B_{s2}  B_1 K$ $T A {\cal P}$ vertices. The results heavily depend on the distribution amplitudes of pseudoscalar mesons. Especially to the parameters of the first and second momenta. However, recent progress on determining the first and second momenta of distribution amplitudes of pseudoscalar mesons within lattice QCD computations enables more precise calculations. The recent values with corresponding uncertainties are taken into account for DAs in this study. We find that the couplings vary widely depending on the Lorentz structures we choose. However, symmetry arguments enable us to choose consistent structures. With the obtained coupling constants, we also estimated the decay widths of the anticipated decays. Even though the decay widths are large, we hope that our findings will be useful for the studies in understanding the nature of heavy axial tensor mesons.
%
%
%
\appendix
\section{Matrix Elements of Non-Local Operators in terms of DAs}
\label{appendix_das}
%
In this section, we present the matrix elements of the non-local operators between the vacuum and one-particle light pseudoscalar meson states in terms of the distribution amplitudes ~\cite{Ball:2006wn,Ball:2004ye}.
\bea
\label{ejly09}
\lla {\cal P}(p)\vel \bar q_1(x) \gamma_\mu \gamma_5 q_1(0)\ver 0 \rra \es
-i f_{\cal P} q_\mu  \int_0^1 du  e^{i \bar u q x}
    \left[ \varphi_{\cal P}(u) + {1\over 16} m_{\cal P}^2
x^2 {\Bbb{A}}(u) \right] \nnb \\
\ek {i\over 2} f_{\cal P} m_{\cal P}^2 {x_\mu\over qx}
\int_0^1 du e^{i \bar u qx} {\Bbb{B}}(u)~,\nnb \\
\lla {\cal P}(p)\vel \bar q_1(x) i \gamma_5 q_2(0)\ver 0 \rra \es
\mu_{\cal P} \int_0^1 du e^{i \bar u qx} \phi_P(u)~,\nnb \\
\lla {\cal P}(p)\vel \bar q_1(x) \sigma_{\alpha \beta} \gamma_5 q_2(0)\ver 0 \rra \es
{i\over 6} \mu_{\cal P} \left( 1 - \widetilde{\mu}_{\cal P}^2 \right)
\left( q_\alpha x_\beta - q_\beta x_\alpha\right)
\int_0^1 du e^{i \bar u qx} \phi_\sigma(u)~,\nnb \\
\lla {\cal P}(p)\vel \bar q_1(x) \sigma_{\mu \nu} \gamma_5 g_s
G_{\alpha \beta}(v x) q_2(0)\ver 0 \rra \es i \mu_{\cal P} \left\{
q_\alpha q_\mu \left[ g_{\nu \beta} - {1\over qx}(q_\nu x_\beta +
q_\beta x_\nu) \right] \right. \nnb \\
\ek q_\alpha q_\nu \left[ g_{\mu \beta} -
{1\over qx}(q_\mu x_\beta + q_\beta x_\mu) \right] \nnb \\
\ek q_\beta q_\mu \left[ g_{\nu \alpha} - {1\over qx}
(q_\nu x_\alpha + q_\alpha x_\nu) \right] \nnb \\
\ar q_\beta q_\nu \left. \left[ g_{\mu \alpha} -
{1\over qx}(q_\mu x_\alpha + q_\alpha x_\mu) \right] \right\} \nnb \\
\cp \int {\cal D} \alpha e^{i (\alpha_{\bar q} +
v \alpha_g) qx} {\cal T}(\alpha_i)~,\nnb \\
\lla {\cal P}(p)\vel \bar q_1(x) \gamma_\mu \gamma_5 g_s
G_{\alpha \beta} (v x) q_2(0)\ver 0 \rra \es q_\mu (q_\alpha x_\beta -
q_\beta x_\alpha) {1\over qx} f_{\cal P} m_{\cal P}^2
\int {\cal D}\alpha e^{i (\alpha_{\bar q} + v \alpha_g) qx}
{\cal A}_\parallel (\alpha_i) \nnb \\
\ar \left\{q_\beta \left[ g_{\mu \alpha} - {1\over qx}
(q_\mu x_\alpha + q_\alpha x_\mu) \right] \right. \nnb \\
\ek q_\alpha \left. \left[g_{\mu \beta}  - {1\over qx}
(q_\mu x_\beta + q_\beta x_\mu) \right] \right\}
f_{\cal P} m_{\cal P}^2 \nnb \\
\cp \int {\cal D}\alpha e^{i (\alpha_{\bar q} + v \alpha _g)
q x} {\cal A}_\perp(\alpha_i)~,\nnb \\
\lla {\cal P}(p)\vel \bar q_1(x) \gamma_\mu i g_s G_{\alpha \beta}
(v x) q_2(0)\ver 0 \rra \es q_\mu (q_\alpha x_\beta - q_\beta x_\alpha)
{1\over qx} f_{\cal P} m_{\cal P}^2 \int {\cal D}\alpha e^{i (\alpha_{\bar q} +
v \alpha_g) qx} {\cal V}_\parallel (\alpha_i) \nnb \\
\ar \left\{q_\beta \left[ g_{\mu \alpha} - {1\over qx}
(q_\mu x_\alpha + q_\alpha x_\mu) \right] \right. \nnb \\
\ek q_\alpha \left. \left[g_{\mu \beta}  - {1\over qx}
(q_\mu x_\beta + q_\beta x_\mu) \right] \right\} f_{\cal P} m_{\cal P}^2 \nnb \\
\cp \int {\cal D}\alpha e^{i (\alpha_{\bar q} +
v \alpha _g) q x} {\cal V}_\perp(\alpha_i)~,\nnb
\eea
where
\bea
\label{nolabel}
\mu_{\cal P} = f_{\cal P} {m_{\cal P}^2\over m_{q_1} + m_{q_2}}~,~~~~~
\widetilde{\mu}_{\cal P} = {m_{q_1} + m_{q_2} \over m_{\cal P}}~, \nnb
\eea
and $q_1$ and $q_2$ are the quarks forming the pseudoscalar meson ${\cal P}$,
${\cal D}\alpha = d\alpha_{\bar q} d\alpha_q d\alpha_g
\delta(1-\alpha_{\bar q} - \alpha_q - \alpha_g)$.
Here 
$\varphi_{\cal P}(u)$ is the leading twist--two, $\phi_P(u)$, $\phi_\sigma(u)$,
${\cal T}(\alpha_i)$ are the twist--three, and
$\Bbb{A}(u)$, $\Bbb{B}(u)$, ${\cal A}_\perp(\alpha_i),$ ${\cal A}_\parallel(\alpha_i),$
${\cal V}_\perp(\alpha_i)$ and ${\cal V}_\parallel(\alpha_i)$
are the twist-four DAs, respectively, whose explicit expressions
are given in Appendix~\ref{twist-four}.
\section{Correlation Functions}
\label{appendix_a}
In this section, we present the expressions of the theoretical part of the correlation functions $\Pi_i^{A(B)}$ and $\Pi_i^{V(B)}$ as well as the coefficients $A_i$ and $B_i$ given in Eq.(\ref{ejly12}). 
%
%
\subsection{Expressions of the coefficients $A_i$ in phenomenological and theoretical parts of the correlation function for $T \to A + \mathcal{P}$ transition}
%
\subsubsection{Coefficients of the  $\epsilon_{\alpha\lambda\rho\tau} p_\lambda q_\rho p_\beta$ and $\epsilon_{\beta\lambda\rho\tau} p_\lambda q_\rho p_\alpha$ structures}
\vspace{-3em}
%
%
\bea
\Pi_1^A \es \Pi_2^A =
  e^{-m_Q^2/M^2}
\Bigg\{
{1\over 48 M^2}f_{\cal P} m_{\cal P}^2 m_Q^2 \mathbb{A}(u_0) - 
{1\over 12} f_{\cal P} M^2 \rho_{\cal P}(u_0) \nnb \\
\ar {1\over 72} \Big[3 f_{\cal P} m_{\cal P}^2 \mathbb{A}(u_0) - 
2 (1 - \widetilde{\mu}_{\cal P}^2) m_Q  \mu_{\cal P}
\phi_{\sigma}(u_0)\Big]
\Bigg\}~, \nnb
\eea
\bea
A_1 = A_2 = -{f_A f_{T} m_A {m_T} 
({m_T}^2 - m_A^2 + m_{\cal P}^2) \over 4}~.\nnb
\eea
Here $\mu^2 = \frac{2 m_T^2 m_{V (A)}^2}{m_T^2 + m_{V (A)}^2}$ and $u_0 = \frac{M_1^2}{M_1^2 + M_2^2}$.
\subsubsection{Coefficients of the $\epsilon_{\alpha\lambda\rho\tau} p_\lambda q_\rho q_\beta$ and $\epsilon_{\beta\lambda\rho\tau} p_\lambda q_\rho q_\alpha$ structures}
\vspace{-3em}
%
%
\bea
\Pi_3^A \es \Pi_4^A =
  e^{-m_Q^2/M^2}
\Bigg\{
{1\over 96 M^2}f_{\cal P} m_{\cal P}^2 m_Q^2 \mathbb{A}(u_0) -
{1\over 24} f_{\cal P} M^2 \rho_{\cal P}(u_0) \nnb \\
\ar {1\over 144} \Big[3 f_{\cal P} m_{\cal P}^2 \mathbb{A}(u_0) -
2 (1 - \widetilde{\mu}_{\cal P}^2) m_Q  \mu_{\cal P}
\phi_{\sigma}(u_0)\Big]
\Bigg\}~, \nnb
\eea
\bea
A_3 = A_4 = {f_A f_{T} m_A {m_T}^3 \over 2}~.\nnb
\eea
%
\subsection{Expressions of the coefficients $B_i$ in phenomenological and theoretical parts of the correlation function for $T \to V + \mathcal{P}$ transition}
%

\subsubsection{Coefficients of the $p_\alpha g_{\beta\tau}$ and  $p_\beta g_{\alpha\tau}$ structures}
\vspace{-3em}
%
%
\bea
\Pi_1^V \es \Pi_2^V =
  e^{-m_Q^2/M^2}
\Bigg\{- {1 \over 48 M^2} f_{\cal P} m_{\cal P}^4 m_Q^2 \mathbb{A}(u_0) \nnb \\
\ek { f_{\cal P} m_{\cal P}^2 \over 96} \Big\{4 m_{\cal P}^2 
\mathbb{A}(u_0) + m_Q^2 \Big[8 \widetilde{j}(\mathbb{B}) -
\mathbb{A}^\prime(u_0)\Big]\Big\} \nnb \\
\ar {M^2\over 48} \Big\{4 m_Q \mu_{\cal P} \phi_{\sigma}(u_0) + 
f_{\cal P} m_{\cal P}^2 \Big[4 \varphi_{\cal P}(u_0) +
      \mathbb{A}^\prime(u_0)\Big]\Big\} - 
{1 \over 24} \Big[f_{\cal P} M^4 \varphi_{\cal P}^\prime(u_0)\Big] \Bigg\}~, \nnb 
\eea
\bea
B_1 = B_2 = {f_{T} f_V {m_T} m_V 
({m_T}^2 - m_V^2 + m_{\cal P}^2) \over 4}~. \nnb 
\eea
\subsubsection{Coefficient of the $p_\tau g_{\alpha\beta}$ structure}
\vspace{-3em}
%
%
%
\bea
\Pi_3^V \es
  e^{-m_Q^2/M^2}
\left[ {1\over 4} f_{\cal P} m_{\cal P}^2 M^2 \widetilde{j}(\mathbb{B})
\right]~,\nnb
\eea
\bea
B_3 \es {f_{T} f_V {m_T} [m_V^4 - 
({m_T}^2 - m_{\cal P}^2)^2)] \over 12 m_V}~. \nnb
\eea
%
%
\subsubsection{Coefficients of the $q_\alpha g_{\beta\tau}$ and $q_\beta g_{\alpha\tau}$ structures}
\vspace{-3em}
%
%
%
\bea
\Pi_4^V \es \Pi_5^V =
  e^{-m_Q^2/M^2}
\Bigg\{
-{1\over 96 M^2} f_{\cal P} m_{\cal P}^4 m_Q^2 \mathbb{A}(u_0) \nnb \\
\ek {1\over 192 } f_{\cal P} m_{\cal P}^2 \Big\{ \Big[2 (2 m_{\cal P}^2 - m_Q^2)
\mathbb{A}(u_0)\Big]
+    m_Q^2 \Big[8 \widetilde{j}(\mathbb{B}) -
\mathbb{A}^\prime(u_0)\Big]\Big\} \nnb \\
\ar  {1\over 288}  M^2 \Big\{ 4 m_Q \mu_{\cal P} \Big[3 \phi_{\sigma}(u_0) - (1 - \widetilde{\mu}_{\cal P}^2)
\phi_{\sigma}(u_0)\Big]\nnb \\
\ar 
    3 f_{\cal P} m_{\cal P}^2 \Big[2 \mathbb{A}(u_0) + 4 \varphi_{\cal P}(u_0) +
\mathbb{A}^\prime(u_0)\Big]\Big\} -
 {1\over 48} f_{\cal P} M^4 \Big[2 \varphi_{\cal P}(u_0) + \varphi_{\cal
P}^\prime(u_0)\Big] \Bigg\}~,\nnb
\eea
\bea
B_4 = B_5 =  -{f_{T} f_V {m_T}^3 m_V \over 2}~. \nnb
\eea
%
\subsubsection{Coefficients of the $q_\tau g_{\alpha\beta}$ and  $q_\beta g_{\alpha\tau}$ structures}
\vspace{-3em}
%
%
%
\bea
\Pi_6^V \es
  e^{-m_Q^2/M^2}
\Bigg\{ {1\over 24}f_{\cal P} m_{\cal P}^2 m_Q^2 \mathbb{A}(u_0) - 
{1\over 6} f_{\cal P} M^4 \varphi_{\cal P}(u_0) \nnb \\
\ar {1\over 144} M^2 \Big\{9 f_{\cal P} m_{\cal P}^2 \Big[\mathbb{A}(u_0) + 2
\widetilde{j}(\mathbb{B})\Big] +
4 (1 - \widetilde{\mu}_{\cal P}^2) m_Q \mu_{\cal P}
\phi_{\sigma}(u_0)\Big\} \Bigg\}~,\nnb
\eea
\bea
B_6 =  {f_{T} f_V {m_T} 
({m_T}^2 + m_V^2 - m_{\cal P}^2)^2 \over 12 m_V}~. \nnb
\eea


\subsubsection{Coefficient of the $p_\alpha p_\beta p_\tau$ structure}
\vspace{-3em}
%
%
\bea
\Pi_7^V \es
  e^{-m_Q^2/M^2}
\Bigg[{1\over 3} f_{\cal P} m_{\cal P}^2 \widetilde{j}(\mathbb{B})
\Bigg]~, \nnb
\eea
\bea
B_7 = - {f_{T} f_V [{m_T}^4 - m_V^4 + 
m_{\cal P}^2 (4 {m_T}^2 + m_{\cal P}^2)] \over 6 {m_T} m_V}~. \nnb
\eea



\subsubsection{Coefficient of the $q_\alpha q_\beta q_\tau$ structure}
\vspace{-3em}
%
%
\bea
\Pi_8^V \es
  e^{-m_Q^2/M^2}
\Bigg\{ {1\over 48 M^2} f_{\cal P} m_{\cal P}^2 m_Q^2 \mathbb{A}(u_0) + 
{1\over 24} f_{\cal P} m_{\cal P}^2 \Big[\mathbb{A}(u_0) 
+ \widetilde{j}(\mathbb{B})\Big] \nnb \\
\ek {1\over 12} f_{\cal P} M^2 \varphi_{\cal P}(u_0) \Bigg\}~,\nnb
\eea
\bea
B_8 = {f_{T} f_V {m_T}^3 \over m_V}~. \nnb
\eea



\subsubsection{Coefficients of the $p_\alpha q_\beta q_\tau$ and  $p_\beta q_\alpha q_\tau$ structures}
\vspace{-3em}
%
%
\bea
\Pi_9^V \es \Pi_{10}^V =
  e^{-m_Q^2/M^2}
\Bigg\{ {1\over 32 M^2}f_{\cal P} m_{\cal P}^2 m_Q^2 \mathbb{A}(u_0)
-{1\over 8} f_{\cal P} M^2 \varphi_{\cal P}(u_0) \nnb \\
\ar {1\over 144} \Big\{3 f_{\cal P} m_{\cal P}^2 \Big[3 \mathbb{A}(u_0) + 
4 \widetilde{j}(\mathbb{B})\Big] + 2 (1 - \widetilde{\mu}_{\cal P}^2) m_Q \mu_{\cal P}
    \phi_{\sigma}(u_0)\Big\} \Bigg\}~,\nnb
\eea
\bea
B_9 = B_{10} = - {f_{T} f_V {m_T} 
({m_T}^2 - m_V^2 + m_{\cal P}^2) \over 2 m_V}~. \nnb
\eea



\subsubsection{Coefficient of the $p_\tau q_\alpha q_\beta$ structure}
\vspace{-3em}
%
%
\bea
\Pi_{11}^V \es
  e^{-m_Q^2/M^2}
\Bigg\{ {1\over 48 M^2}f_{\cal P} m_{\cal P}^2 m_Q^2 \mathbb{A}(u_0)
 - {1\over 12} f_{\cal P} M^2 \varphi_{\cal P}(u_0) \nnb \\
\ar {1\over 72} \Big\{3 f_{\cal P} m_{\cal P}^2 \Big[\mathbb{A}(u_0) + 
2 \widetilde{j}(\mathbb{B})\Big] - 2 (1 - \widetilde{\mu}_{\cal P}^2) m_Q \mu_{\cal P}
    \phi_{\sigma}(u_0)\Big\} \Bigg\}~,\nnb
\eea
\bea
B_{11} =  -{f_{T} f_V {m_T}^3 \over m_V}~.\nnb
\eea


\subsubsection{Coefficient of the $p_\alpha p_\beta q_\tau$ structure}
\vspace{-3em}
%
%
\bea
\Pi_{12}^V \es
  e^{-m_Q^2/M^2}
\Bigg\{ {1\over 24 M^2}f_{\cal P} m_{\cal P}^2 m_Q^2 \mathbb{A}(u_0)
 - {1\over 6} f_{\cal P} M^2 \varphi_{\cal P}(u_0) \nnb \\
\ar {1\over 36} \Big\{3 f_{\cal P} m_{\cal P}^2 \Big[\mathbb{A}(u_0) +
2 \widetilde{j}(\mathbb{B})\Big] + 2 (1 - \widetilde{\mu}_{\cal P}^2) m_Q \mu_{\cal P}
    \phi_{\sigma}(u_0)\Big\} \Bigg\}~,\nnb
\eea
\bea
B_{12} =  {f_{T} f_V [{m_T}^4 + (m_V^2 - m_{\cal P}^2)^2 - 
4 {m_T}^2 (m_V^2 - m_{\cal P}^2)] \over 6 {m_T} m_V}~. \nnb
\eea

\subsubsection{Coefficients of the $p_\alpha p_\tau q_\beta$ and  $p_\beta p_\tau q_\alpha$ structures}
\vspace{-3em}
%
%
\bea
\Pi_{13}^V \es \Pi_{14}^V =
  e^{-m_Q^2/M^2}
\Bigg\{ {1\over 24 M^2}f_{\cal P} m_{\cal P}^2 m_Q^2 \mathbb{A}(u_0)
 - {1\over 6} f_{\cal P} M^2 \varphi_{\cal P}(u_0) \nnb \\
\ar {1\over 36} \Big\{3 f_{\cal P} m_{\cal P}^2 \Big[\mathbb{A}(u_0) +
4 \widetilde{j}(\mathbb{B})\Big] + 2 (1 - \widetilde{\mu}_{\cal P}^2) m_Q \mu_{\cal P}
    \phi_{\sigma}(u_0)\Big\} \Bigg\}~,\nnb
\eea
\bea
B_{13} = A_{14} = {f_{T} f_V {m_T} 
({m_T}^2 + m_{\cal P}^2) \over 2 m_V}~.\nnb
\eea
The function $\tilde{j}(f(u))$ is defined as
\begin{equation}
  \label{eq:2}
  \tilde{j}(f(u)) = \int_{u_0}^{1} du f(u).
\end{equation}
Note that the continuum subtraction is taken into account via $M^2 \to M^2(1-e^{-(s_0 - m_Q^2)/M^2})$ and $M^4 \to M^4(1 + \frac{m_Q^2}{M^2} - (1 + \frac{s_0}{M^2})) e^{-(s_0 - m_Q^2)/M^2}$. 

%
\section{Expressions of Distribution Amplitudes for Pseudoscalar Mesons}
\label{twist-four}

In this section, we present the correlation functions of the pseudoscalar mesons in terms of DAs.
\bea
\label{ejly17}
\varphi_{\cal P}(u) \es 6 u \bar u \left[ 1 + a_1^{\cal P} C_1(2 u -1) +
a_2^{\cal P} C_2^{3/2}(2 u - 1) \right]~,  \nnb \\
{\cal T}(\alpha_i) \es 360 \eta_3 \alpha_{\bar q} \alpha_q
\alpha_g^2 \left[ 1 + w_3 {1\over 2} (7 \alpha_g-3) \right]~, \nnb \\
\phi_P(u) \es 1 + \left[ 30 \eta_3 - {5\over 2}
{1\over \mu_{\cal P}^2}\right] C_2^{1/2}(2 u - 1)  \nnb \\
\ar \left( -3 \eta_3 w_3  - {27\over 20} {1\over \mu_{\cal P}^2} -
{81\over 10} {1\over \mu_{\cal P}^2} a_2^{\cal P} \right)
C_4^{1/2}(2u-1)~, \nnb \\
\phi_\sigma(u) \es 6 u \bar u \left[ 1 + \left(5 \eta_3 - {1\over 2} \eta_3 w_3 -
{7\over 20}  \mu_{\cal P}^2 - {3\over 5} \mu_{\cal P}^2 a_2^{\cal P} \right)
C_2^{3/2}(2u-1) \right] ~, \nnb \\
{\cal V}_\parallel(\alpha_i) \es 120 \alpha_q \alpha_{\bar q} \alpha_g
\left( v_{00} + v_{10} (3 \alpha_g -1) \right) ~, \nnb \\
{\cal A}_\parallel(\alpha_i) \es 120 \alpha_q \alpha_{\bar q} \alpha_g
\left( 0 + a_{10} (\alpha_q - \alpha_{\bar q}) \right) ~, \nnb \\
{\cal V}_\perp (\alpha_i) \es - 30 \alpha_g^2\left[ h_{00}(1-\alpha_g) +
h_{01} (\alpha_g(1-\alpha_g)- 6 \alpha_q \alpha_{\bar q}) +
h_{10}(\alpha_g(1-\alpha_g) - {3\over 2} (\alpha_{\bar q}^2+
\alpha_q^2)) \right] ~, \nnb \\
{\cal A}_\perp (\alpha_i) \es 30 \alpha_g^2(\alpha_{\bar q} - \alpha_q)
\left[ h_{00} + h_{01} \alpha_g + {1\over 2} h_{10}(5 \alpha_g-3) \right] ~, \nnb \\
\Bbb{B}(u) \es g_{\cal P}(u) - \varphi_{\cal P}(u) ~, \nnb \\
g_{\cal P}(u) \es g_0 C_0^{1/2}(2 u - 1) + g_2 C_2^{1/2}(2 u - 1) +
g_4 C_4^{1/2}(2 u - 1) ~, \nnb \\
\Bbb{A}(u) \es 6 u \bar u \left[{16\over 15} + {24\over 35} a_2^{\cal P}+
20 \eta_3 + {20\over 9} \eta_4 +
\left( - {1\over 15}+ {1\over 16}- {7\over 27}\eta_3 w_3 -
{10\over 27} \eta_4 \right) C_2^{3/2}(2 u - 1)  \right. \nnb \\
    \ar \left. \left( - {11\over 210}a_2^{\cal P} - {4\over 135}
\eta_3w_3 \right)C_4^{3/2}(2 u - 1)\right] ~, \nnb \\
\ar \left( -{18\over 5} a_2^{\cal P} + 21 \eta_4 w_4 \right)
\left[ 2 u^3 (10 - 15 u + 6 u^2) \ln u  \right. \nnb \\
\ar \left. 2 \bar u^3 (10 - 15 \bar u + 6 \bar u ^2) \ln\bar u +
u \bar u (2 + 13 u \bar u) \right]~, \nnb
\eea
where $C_n^k(x)$ are the Gegenbauer polynomials, and
\bea
\label{ejly18}
h_{00}\es v_{00} = - {1\over 3}\eta_4 ~, \nnb \\
a_{10} \es {21\over 8} \eta_4 w_4 - {9\over 20} a_2^{\cal P} ~, \nnb \\
v_{10} \es {21\over 8} \eta_4 w_4 ~, \nnb \\
h_{01} \es {7\over 4}  \eta_4 w_4  - {3\over 20} a_2^{\cal P} ~, \nnb \\
h_{10} \es {7\over 4} \eta_4 w_4 + {3\over 20} a_2^{\cal P} ~, \nnb \\
g_0 \es 1 ~, \nnb \\
g_2 \es 1 + {18\over 7} a_2^{\cal P} + 60 \eta_3  + {20\over 3} \eta_4 ~, \nnb \\
g_4 \es  - {9\over 28} a_2^{\cal P} - 6 \eta_3 w_3~, \nnb \\
\eta_3 \es 0.015~, \nnb \\
\eta_4 \es 10~, \nnb  \\
w_3 \es -3~,  \nnb  \\
w_4 \es 0.1. \nnb 
\eea
%
%
%


\bibliographystyle{utcaps_mod}
\bibliography{all.bib}



\end{document}